\documentclass[acus]{JAC2003}
\usepackage{url}
\usepackage{graphicx}
\begin{document}
\title{A COMPLETE SCHEME FOR A MUON COLLIDER~\thanks{This work was presented by A.~Sessler at the workshop. This work was supported by US Department of Energy under contracts AC02-98CH10886 and DE-AC02-76CH03000}}

\author{Robert B.~Palmer, J. Scott~Berg, Richard C.~Fernow, Juan Carlos~Gallardo, Harold G.~Kirk\\
 (BNL, Upton, NY); Yuri~Alexahin,  David~Neuffer  (Fermilab, Batavia, IL);
 Stephen Alan~Kahn\\ (Muons Inc, Batavia, IL);
 Don J.~Summers (University of Mississippi, Oxford, MS)}
\maketitle
\vskip.5in
\begin{abstract}
A complete scheme for production, cooling, acceleration, and ring for a
1.5~TeV center of mass muon collider is presented, together with parameters for two higher energy machines.  The schemes starts with the front end of a proposed neutrino factory that yields bunch trains of both muon signs. Six dimensional cooling in long-period helical lattices reduces the longitudinal emittance until it becomes possible to merge the trains into single bunches, one of each sign. Further cooling in all dimensions is applied to the single bunches in further helical lattices. Final transverse cooling to the required parameters is achieved in 50~T solenoids.
\end{abstract}

\begin{table}[hbt]
\begin{center}
\caption{Parameters of three muon colliders.
 Note 1: Depth is relative to any nearby low land, e.g. Fox river at FNAL. Note 2: Survival is from the end of phase rotation to the collider ring.}
\begin{tabular}{|l|ccc|}
\hline
  $E_{\rm c.m.s}  $ ~~~~~~~~~~~(TeV)& 1.5&  4& 8\\
 $\cal{L}$~~~~~($10^{34}$ cm$^2$sec$^{-1}$) &  1 &  4  & 8  \\
\hline
Beam-beam $\Delta\nu$ & 0.1& 0.1&0.1\\
$\mu$/bunch~~~~~~~~~~($10^{12}$)& 2& 2& 2\\
 $<B_{\rm ring}>$~~~~~~~~~~~(T)& 5.2& 5.2&10.4\\
$\beta^*  = \sigma_z$~~~~~~~~~~(mm)& 10&3&3\\
rms $dp/p$~~~~~~~~~~~~~(\%)& 0.09&0.12&0.06\\

Depth for $\nu$ rad~$^1$~(m)& 13 & 135 & 540\\
Muon Survival~$^2$& $\approx$0.07 & $\approx$0.07& $\approx$0.07\\
\hline
 Rep. rate~~~~~~~~~~~~~(Hz)& 13  & 6&3\\
 $P_{\rm driver}$~~~~~~~~~~~(MW)& $\approx$4 & $\approx$ 1.8& $\approx$ 0.8\\
 $\epsilon_\perp$~~~~~~~($\pi$ mm mrad)& 25 &  25& 25\\
 $\epsilon_\parallel$~~~~~~~~~($\pi$ mm rad) & 72    & 72& 72\\
\hline
\end{tabular}
\label{parameters}
\end{center}
\end{table}

\begin{figure}[htb]
\centering
\includegraphics*[width=75mm]{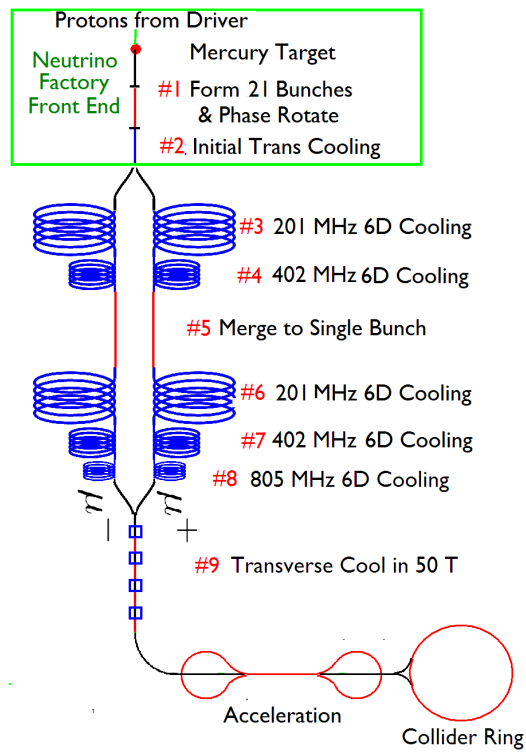}
\caption{(Color) Schematic of the components of a Muon Collider.}
\label{schematic}
\end{figure}

\begin{figure}[htb]
\centering
\includegraphics*[width=75mm]{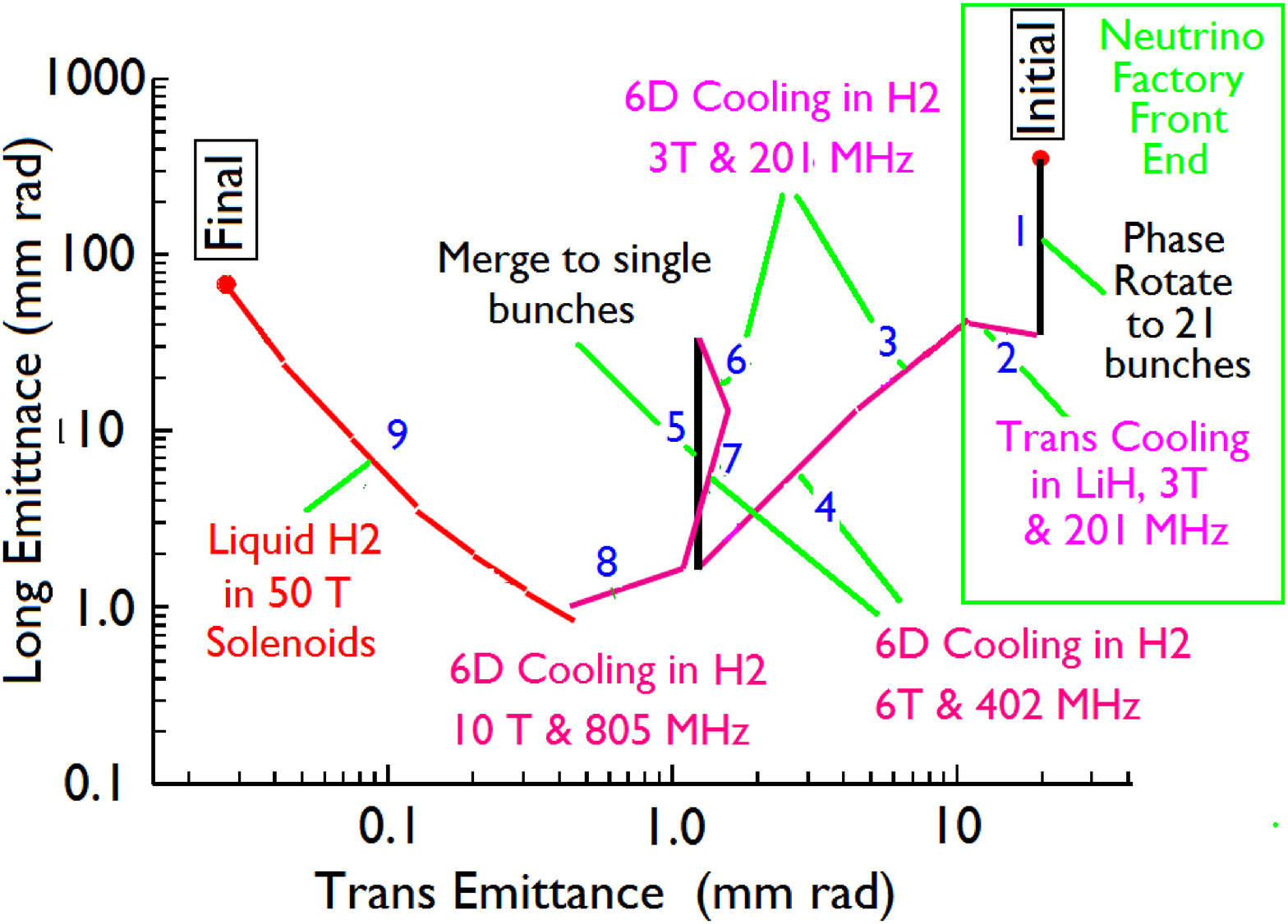}
\caption{(Color) Transverse vs. longitudinal emittances before and after each
  stage. The nine stages are indicated with the numeral 1--9.}
\label{longtrans}
\end{figure}
\section{INTRODUCTION}

Muon colliders were first proposed by Budker in 1969~\cite{budker}, and later discussed by others~\cite{skrinsky}. A more detailed study was done for Snowmass 96~\cite{feasibility}, but in none of these was a complete scheme defined for the manipulation and cooling of the required muons.  

Muon colliders would allow the high energy study of point-like collisions of leptons without some of the difficulties associated with high energy electrons; e.g. the synchrotron radiation requiring their acceleration to be essentially linear, and as a result, long.  Muons can be accelerated in smaller rings and offer other advantages, but they are produced only diffusely and they decay rapidly, making the detailed design of such machines difficult.  In this paper, we outline a complete scheme for capture, phase manipulation and cooling of the muons, every component of which has been simulated at some level.

The work in this paper was performed as part of the NFMCC collaboration~\cite{nfmcc}, the recently formed MCTF~\cite{mctf}, and Muons Inc.~\cite{muonsinc}.
\section{COLLIDER PARAMETERS}
Table ~\ref{parameters} gives parameters for muon colliders at three energies.  Those at 1.5~TeV  correspond to a recent collider ring design~\cite{alexahin}. The 4~TeV example is taken from the 96--study~\cite{feasibility}.  The 8~TeV is an extrapolation assuming higher  bending fields and more challenging interaction point parameters. All three use the same muon intensities and emittances, although the repetition rates for the higher energy machines are reduced to control neutrino radiation.
\section{PROPOSED SYSTEM}
Figure~\ref{schematic} shows a schematic of the components of the  system. Figure~\ref{longtrans} shows a plot of the longitudinal and transverse emittances of the muons as they progress from production to the specified requirements for the colliders.  The subsystems used to manipulate and cool the beams to meet these requirements are indicated by the numerals 1--9 on the figures.
\subsection{Proton Driver}
The proton driver requirements depend on the muon survival estimates that
will be discussed in a later section. We further assume, from the neutrino
factory studies, that pion production in the 21~best bunches, at the end of
phase rotation, is 1.7\% per proton per GeV. The resulting required proton
bunches, for different energies, are given in Tb.~\ref{pspace}. For
efficiency in the following phase rotation, an rms bunch length of 3~ns is
required. The space charge tune shift and required longitudinal phase space
densities are challenging at the lower proton energies, but easier at the
higher energies.
\begin{table}[hbt]
\begin{center}
\caption{Proton bunch intensity for three different proton energies}
\begin{tabular}{|lccc|}
\hline
 E  (GeV)&8&24&60\\
 N$_p$~(10$^{13}$)&21&7&2.8\\
\hline
\end{tabular}
\label{pspace}
\end{center}
\end{table}
\subsection{Production, Phase Rotation, and Initial Cooling}
The muons are generated by the decay of pions produced by proton bunches interacting in a mercury jet target~\cite{merit}.  These pions are captured by a 20~T solenoid surrounding the target, followed by an adiabatic lowering of the field to 3~T in a decay channel.  

\begin{figure}[htb]
\centering
\includegraphics*[width=75mm]{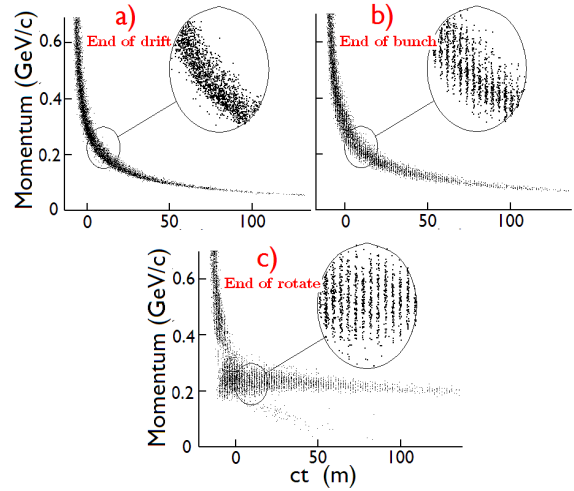}
\caption{Phase spaces during phase rotation a) before bunching,  b) after bunching,   c) after rotation.}
\label{rotate}
\end{figure}

The first manipulation (\#1), referred to as phase
rotation~\cite{neuffer-rot}, converts the initial single short muon bunch
with very large energy spread into a train of bunches with much reduced
energy spread of which we use only 21.  The initial bunch is allowed to lengthen and develop a
time-energy correlation in a 110~m drift.  It is then bunched into a train,
without changing the time-energy correlation, using rf cavities whose
frequency varies with location, falling from 333~MHz to 234~MHz.  Then, by
phase and frequency control, the rf accelerates the low energy bunches and
decelerates the high energy ones. Figure~\ref{rotate} shows 
ICOOL~\cite{icool} simulations of the phase spaces before bunching, after bunching, and after rotation.

 Muons of both signs are captured and then (\#2) cooled transversely in a linear channel using LiH absorbers, periodic alternating 2.8~T solenoids, and pillbox 201~MHz rf cavities.  All the components up to this point are identical to those described in a recent study~\cite{study2a} for a Neutrino Factory.

\begin{figure}[htb]
\centering
\includegraphics*[width=75mm]{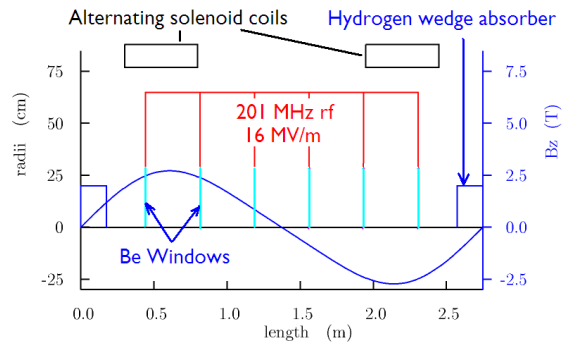}
\caption{(Color) One cell of the first 6D cooling lattice.}
\label{rfofo-cell1}
\end{figure}

\subsection{6D Cooling Before Merge}
\begin{figure}[htb]
\centering
\includegraphics*[width=70mm]{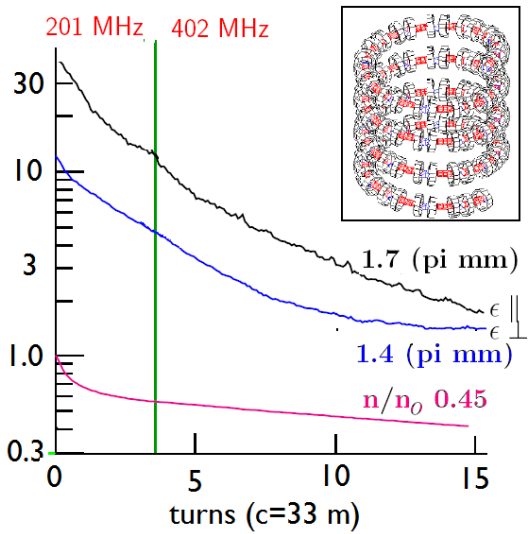}
\caption{(Color) ICOOL simulation of 6D cooling in stages (\#3) \& (\#4). Inset: long pitch helical geometry of (\#3).}
\label{rfofo34}
\end{figure}

The next stage (\#3) cools simultaneously in all 6 dimensions.  The
lattice~\cite{rfofo}, shown in figure~\ref{rfofo-cell1}, uses 3~T solenoids
for focus, weak dipoles (generated by tilting the solenoids) to generate
bending and dispersion, wedge shaped liquid hydrogen filled absorbers where
the cooling takes place, and 201~MHz rf, to replenish the energy lost in
the absorbers.  The dipole fields cause the lattices to curve, forming a
slow upward or downward helix (see inset in Fig.~\ref{rfofo34}). 
The following stage (\#4) uses a lattice essentially the same as (\#3), but
with twice the field strength, half the geometric dimensions, and 402
instead of 201~MHz rf.  Figure~\ref{rfofo34} shows the results of a
simulation of both systems using ICOOL.  Although this simulation was done
for circular, rather than the helical geometry, it used realistic coil and rf geometries.  Preliminary studies~\cite{guggenheim} suggest that the differences introduced by the helical, instead of circular, geometries will be negligible. The simulation did not include the required matching between the two stages. The simulations also used fields that, while they satisfied Maxwell's equations and had realistic strengths, were not actually calculated from specified coils. Simulations reported in reference~\cite{rfofo}, using fields from actual coils, gave slightly better results.

\begin{figure}[htb]
\centering
\includegraphics*[width=\linewidth]{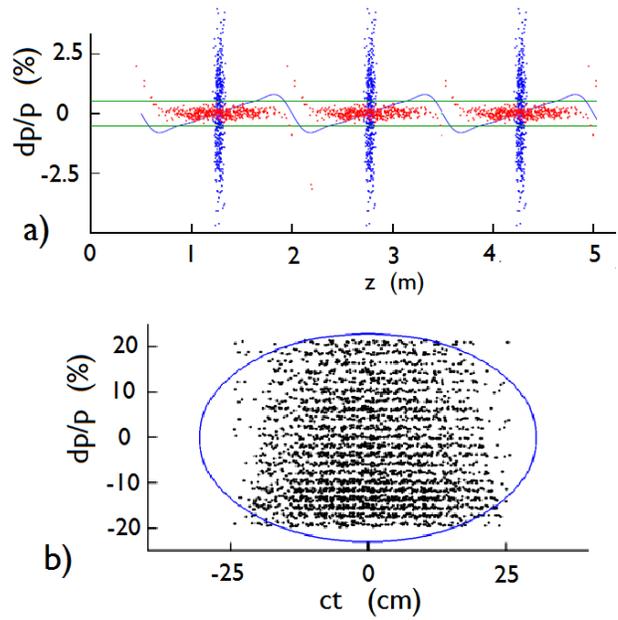}
\caption{(Color) 1D Simulation of merge (\#5): a) before (blue) and after (red) first rotation, b) after second rotation.}
\label{merge}
\end{figure}

\subsection{Bunch Merge}
Since collider luminosity is proportional to the square of the number of muons per bunch, it is important to use relatively few bunches with many muons per bunch. Capturing the initial muon phase space directly into single bunches requires low frequency ($\approx$ 30~MHz) rf, and thus low gradients, resulting in slow initial cooling. It is thus advantageous to capture initially into multiple bunches at 201~MHz and merge them after cooling allows them to be recombined into a single bunch at that frequency.  This recombination (\#5) is done in two stages:  a) using a drift followed by 201~MHz rf, with harmonics, the individual bunches are phase rotated to fill the spaces between bunches and lower their energy spread; followed by b) 5~MHz rf, plus harmonics, interspersed along a long drift to phase rotate the train into a single bunch that can be captured using 201~MHz.  Results of an initial one dimensional simulation of this process is shown in Fig.~\ref{merge}.  Work is ongoing on the design and simulation of a system with the low frequency rf separated from a following drift in a wiggler system with negative momentum compaction to reduce the length and decay losses.

\subsection{6D Cooling After Merge}
\begin{figure}[htb]
\centering
\includegraphics*[width=75mm]{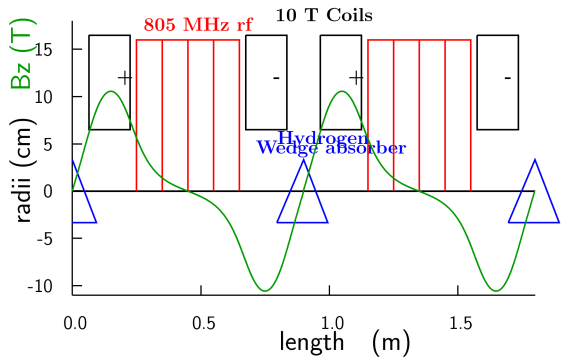}
\caption{(Color) One cell of the last 6D cooling lattice.}
\label{rfofo-cell2}
\end{figure}

\begin{figure}[tb]
\centering
\includegraphics*[width=75mm]{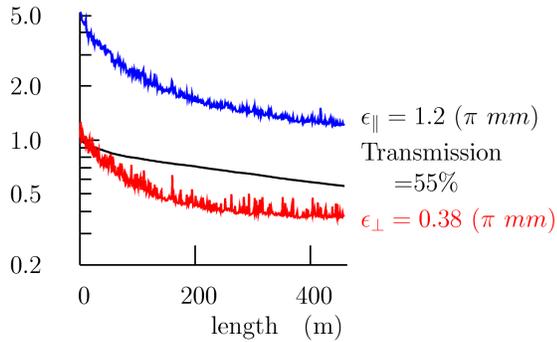}
\caption{(Color) ICOOL simulation of final 6D cooling lattice (\#8)
 using 10~T solenoids and 805~MHz rf.}
\label{last6d}
\end{figure}

After the bunch merging, the longitudinal emittance of the single bunch is
now similar to that at the start of cooling.  It can thus be taken through
the same, or similar, cooling systems as (\#3) and (\#4):  now numbered
(\#6) and (\#7).  One more (\#8) stage of 6D cooling has been designed
(Fig.~\ref{rfofo-cell2}), using 10~T magnets, hydrogen wedge absorbers, and
805~MHz rf. Its ICOOL simulated performance is show in Fig.~\ref{last6d}.
Again, the simulation shown used fields that, while they satisfied
Maxwell's equations and had realistic strengths, were not actually
calculated from specified coil configuration.

\begin{figure}[tb]
\centering
\includegraphics*[width=65mm]{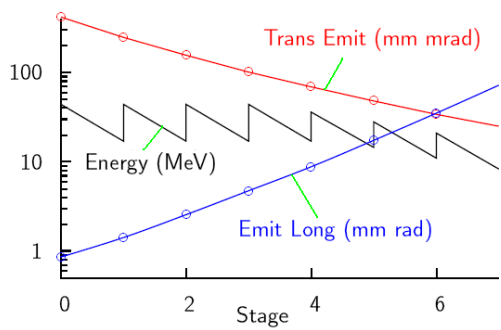}
\caption{(Color) Results of ICOOL simulations of transverse cooling in liquid hydrogen in 7 sequential 50~T solenoids.}
\label{50T}
\end{figure}

\subsection{Final Transverse Cooling in High Field Solenoids}
To attain the required final  transverse emittance, the cooling needs stronger focusing than is achievable in the 6D cooling lattices used in the earlier stages. It can be obtained in liquid hydrogen in strong solenoids, if the momentum is allowed to fall sufficiently low. But at the lower momenta the momentum spread, and thus longitudinal emittance, rises relatively rapidly. However, as we see from Fig.~\ref{longtrans}, the longitudinal emittance after (\#8) is far less than that required, so such a rise is acceptable. Figure~\ref{50T} shows the results of ICOOL simulation of cooling in seven 50~T solenoids. The simulation did not include the required matching and re-accelerations between the solenoids.

A 45~T hybrid Cu and superconductor solenoid\cite{45T} is currently
operating at NHFML and an upgrade to 50~T is planned, but this magnet uses
a lot of power. A 27~T solenoid using an 8~T YBCO insert at 4K has recently
been successfully tested~\cite{27T}. There is a conceptual design of an all
superconducting 50~T solenoid~\cite{50T}.

\subsection{Acceleration}
 Sufficiently rapid acceleration is straightforward in linacs and recirculating linear accelerators (RLAs). 
 Lower cost solutions might use fixed field alternating gradient (FFAG) accelerators, rapidly pulsed magnet synchrotrons, and/or
 hybrid SC and pulsed magnet synchrotrons\cite{summers}. 

\subsection{Muon Losses}
The estimate of muon losses is very preliminary. The simulations assumed Gaussian initial distributions and were not very well matched into each lattice, leading to larger initial losses. And no tapering of the focus parameters as function of length was included, leading to larger losses as the emittances approached their equilibrium. As a result, the losses observed were larger than those deduced using the performances in the mid range of each simulation. Table~\ref{losses} shows the result of an attempt to estimate realistic losses, but this remains very preliminary. Since it is this estimate that was used to determine the required proton driver specifications used above, these too must be considered very preliminary.

\begin{table}[hbt]
\begin{center}
\caption{Calculated transmission tune shifts at different stages in the system.}
\begin{tabular}{|lcc|}
\hline
 &Transmission&Cumulative\\
\hline
Linear transverse pre-cooling&0.7&0.7\\
Pre-merge RFOFO cooling&$\approx$ 0.5&0.35\\
Merging &0.8&0.28\\
Post-merge RFOFO cooling&$\approx$ 0.5&0.14\\
Final 50~T solenoid cooling&0.7&0.1\\
Acceleration to 0.75~TeV&0.7&0.07\\
\hline
\end{tabular}
\label{losses}
\end{center}
\end{table}

\subsection{Collider Ring}
For the 1.5~TeV c.m.s, a lattice has been developed~\cite{alexahin}. The parameters as given in Tb.~\ref{parameters} were achieved with adequate momentum acceptance but with dynamic transverse acceptance of only at little over $2\sigma$ for the specified final emittance. We note however that since luminosity is dependent on the square of the bunch densities, there would be little luminosity loss if the larger amplitudes were collimated prior to injection into the ring. 

\subsection{Space Charge Tune Shifts}
For bunches with Gaussian distributions in all dimensions:
$${\Delta \nu \over \nu_{\rm cell}}
~=~\left({N_\mu\over  \epsilon_\perp  }\right)~
{\beta_{\perp~\rm ave}~ r_\mu \over 2 \sqrt{2\pi}\sigma_z~ ~\beta_v\gamma^2} $$

where $\beta_{\perp~\rm ave}= \left( {L_{\rm cell}\over 2\pi~ \nu_{\rm cell}} \right)$
 and $r_\mu=1.35~10^{-17}~{\rm m} $

Then at the the ends of a number of stages in this system, one obtains the tune shifts given in Tb.~\ref{tuneshift}.
\begin{table}[hbt]
\begin{center}
\caption{Calculated maximum space charge tune shifts at different stages in the system.}
{\small
\begin{tabular}{|ccccccc|}
\hline
 & $N_{\mu}$  & $\beta_{\perp \rm ave}$ & $\sigma_z$ &  $\epsilon_\perp$ & p  &  $\Delta\nu/\nu$ \\
 & $10^{12}$  & mm & mm&  $\pi~\mu$m  &  MeV/c &  \% \\
\hline
(\#4)  &   2 & 292 &  27 &  1500 & 200 &    0.9  \\
(\#6)  &  12 & 584 & 199 &  1500 & 200 &    1.6  \\
(\#7)  &   9 & 292 &  20 &  1500 & 200 &    5.9  \\
(\#8)   &   6 & 191 &  13 &   400 & 200 &   14.5  \\
(\#9$_1$)  &   6 & 222 &  27 &   400 & 100 &   26.1  \\
(\#9$_7$)  &   3 &  93 & 354 &    25 &  42 &   20.0  \\
\hline
\end{tabular}
}
\label{tuneshift}
\end{center}
\end{table}

Note that $N_\mu$ is larger at earlier cooling stages to allow for losses. The first order shifts can be corrected by increasing the focus strength, but tune spreads of half the shifts cannot be corrected.

Before the merge, the shifts are small because the  numbers of muons per bunch are small. The only  6D cooling stage with significant tune shift is the last (\#8). Its tune accepted $\Delta\nu/\nu\approx 0.7$  which is 5 times the calculated maximum full tune spread of $\pm$ 7.3\%, and is not expected to be a problem.

The tune shifts in the 50~T cooling will be significant only during the reaccelerations, where we have assumed $\beta_\perp$s corresponding to 3~T focusing fields. The design of these lattices to accept such tune shifts appears possible, although we are clearly nearing the limit.

\section{ONGOING STUDIES}
There is a serious question as to whether the specified gradients of rf cavities operating under vacuum would operate in the specified magnetic fields. This is under study by NFMCC collaboration~\cite{nfmcc} and alternative designs using high pressure hydrogen gas, or open cell rf with solenoids in the irises, are being considered.  The bunching and phase rotation will be optimized for the muon collider, instead of being copied from a neutrino factory.  Instead of the slow helices, a planar wiggler lattice is being studied that would cool both muon signs simultaneously, thus greatly simplifying the system. The use of more, but lower field (e.g., 35~T) final cooling solenoids is also under study. Experiments are underway to demonstrate two of the new technologies: mercury target~\cite{merit}, ionization cooling~\cite{mice}. Further experimental studies are needed.

\section{CONCLUSION}
Although much work remains to be done, the scenario outlined here appears to be a plausible solution to the problems of capturing, manipulating, and cooling muons to the specifications for muon colliders with useful luminosities and energies, even up to 8~TeV in the center of mass.


\begin{thebibliography}{99}
\bibitem{budker}G. I. Budker, quoted and translated in~\cite{aip352:cline}.
\bibitem{aip352:cline}
  D. Cline, in \textit{Physics Potential and Development of $\mu^+\mu^-$
    Colliders: Second Workshop; Sausalito, CA  1994}, edited by D.~Cline
  (American Institute of Physics, Woodbury, NY, 1996),
  pp.~3--6.
\bibitem{skrinsky}A. N. Skrinsky, quoted and translated in~\cite{aip352:cline}; V. V. Parkhomchuk and A. N. Skrinsky, in \textit{Proceedings of the 12th International Conference on High Energy Accelerators, 1983}, edited by F. T. Cole and R. Donaldson, 485; 
D.~Neuffer, Fermilab report FN-319, 1979;  D.~Neuffer, Part.\ Accel.\ \textbf{14}, 75 (1983);  D.~V.~Neuffer and R.~B.~Palmer, in \textit{Fourth European Particle Accelerator Conference}, edited by V. Suller and Ch.~Petit-Jean-Genaz (World Scientific, Singapore, 1994), p.~52.





\bibitem{feasibility}The $\mu^+\mu^-$ Collaboration, reports
  BNL-52503, Fermi Lab-Conf. 96/092, LBNL-38946, 1996.
\bibitem{nfmcc}http://www.cap.bnl.gov/mumu
\bibitem{mctf}http://mctf.fnal.gov
\bibitem{muonsinc}http://www.muonsinc.com
\bibitem{merit}H. G.~Kirk~\textit{et al.}, \textit{A High-Power Target
    Experiment at the CERN PS} PAC07, Albuquerque, NM, (MOPAS094).
\bibitem{alexahin}Y.~Alexahin and E.~Gianfelice-Wendt,
  Fermilab report Beams-doc-2724-v1, 2007.
  URL: http://beamdocs.fnal.gov/\break AD\mbox{-}public/DocDB/DocumentDatabase
\bibitem{neuffer-rot}D.~Neuffer, report NFMCC-doc-269-v1, 2003.
  URL: http://nfmcc-docdb.fnal.gov/cgi-bin/DocumentDatabase/
\bibitem{icool} ICOOL, http://pubweb.bnl.gov/people/fernow/icool/readme.html
\bibitem{study2a}J. S.~Berg \textit{et al.},
  Phys. Rev. ST Accel. Beams \textbf{9}, 011001 (2006).
\bibitem{rfofo}
  R.~Palmer \textit{et al.}, Phys.\ Rev.\ ST Accel.\ Beams \textbf{8},
  061003 (2005).
\bibitem{guggenheim}
  Amit Klier, presentation at the
  Low Emittance Muon Collider Workshop, Fermilab (6-10 Feb 2006).
  URL: http://www.muonsinc.com/mcwfeb06/
\bibitem{45T}G. B.~Lubkin, Phys. Today 47N12, 21 (1994).
\bibitem{27T}http://www.magnet.fsu.edu/mediacenter/news/pressreleases/2007august7.html; http://www.superpower-inc.com/20070807.aspx
\bibitem{50T}S. A.~Kahn~\textit{et al.}, \textit{A High Field HTS Solenoid
    for Muon Cooling} PAC07, Albuquerque NM, (MOPAN118).
\bibitem{summers}D.J.~Summers \textit{et al.}, \textit{Muon Acceleration to
    750~GeV in the Tevatron tunnel for a 1.5~TeV $\mu^+--\mu^-$ Collider},
  PAC07, Albuquerque NM, (THPMS082); also arXiv:0707.0302.
\bibitem{mice}M.~Zisman, \textit{Status of the International Muon
    Ionization Cooling Experiment}, PAC07, Albuquerque NM, (THPMN119);
  P.~Drumm, \textit{MICE: The International Muon Ionisation Cooling
    experiment}, PAC07, Albuquerque NM, (ROAA004). 
\end{thebibliography}
\end{document}